\documentclass[doublecol,a4paper,showpacs]{epl2} 
\usepackage{tensor}
\usepackage{graphicx}
\usepackage{amsmath}
\usepackage{amssymb}
\usepackage{enumerate}
\usepackage{subfigure}
\usepackage{tabularx}
\usepackage[colorlinks=true, pdfstartview=FitV, linkcolor=blue, citecolor=red, urlcolor=black, breaklinks=true]{hyperref}
\newcommand{\be}{\begin{equation}}
\newcommand{\ee}{\end{equation}}
\newcommand{\ben}{\begin{eqnarray}}
\newcommand{\een}{\end{eqnarray}}
\newcommand{\bes}{\begin{subequations}}
\newcommand{\ees}{\end{subequations}}
\def\bal#1\eal{\begin{align}#1\end{align}}\newcommand{\vphi}{\varphi}

\newcommand{\LL}{{\mathcal L}}

\title{Stable finite energy global vortices and asymptotic freedom}

\author{D. Bazeia\inst{1}\thanks{\email{bazeia@fisica.ufpb.br}} \and M.A. Marques\inst{1}\thanks{Corresponding author; \email{mam@fisica.ufpb.br}} \and R. Menezes\inst{2,1}\thanks{\email{rmenezes@dce.ufpb.br}}}
\shortauthor{D. Bazeia \etal}

\institute{                    
  \inst{1} Departamento de F\'\i sica, Universidade Federal da Para\'\i ba, 58051-970 Jo\~ao Pessoa, PB, Brazil\\
  \inst{2} Departamento de Ci\^encias Exatas, Universidade Federal da Para\'{\i}ba, 58297-000 Rio Tinto, PB, Brazil}


\abstract{This work deals with global vortices in the three-dimensional spacetime. We study the case of a simple model with $U(1)$ symmetry and find a way to describe stable, finite energy global vortices. The price we pay to stabilize the solution is the presence of scale invariance, but we have found a way to trade it with an electric charge in a medium with generalized permittivity, which is further used to capture the basic feature of asymptotic freedom.}

\begin{document}

\maketitle

\section{Introduction}
Kinks, vortices and monopoles are well-known topological structures that appear in relativistic models in $(1,1)$, $(2,1)$, and $(3,1)$ spacetime dimensions, respectively. They find applications in a diversity of contexts of current interest in high energy physics \cite{Col,VS,MS} and in condensed matter \cite{cm1,cm2}. These topological structures have been used to help us understand several phenomena that appear in nonlinear science; see, e.g., Refs.~\cite{app01,app02,app03,app04,app1,app2,app3,app4,app5,app6} and references therein for the use of kinks, vortices and monopoles in a diversity of problems in physics. In particular, in \cite{app01} it is experimentally verified that the presence of electric pulses may change the polarity of a kinklike structure, in \cite{app02} the authors describe kinks in a massive nonlinear sigma model that are related to the long wavelength limit of certain ferromagnetic spin chains, in \cite{app04} it is observed the presence of vortices with very long lifetime in an atomic Bose-Einstein condensate after free expansion, in \cite{app3} the investigation describe new solutions for non-Abelian cosmic strings, in \cite{app5} the authors deal with vortices and monopoles in Weil metals with Fermi surfaces with nontrivial topology, and in \cite{app6} the results show the existence of fermionic zero modes in the background of a Kaluza-Klein monopole, providing an explanation for the decoupling of the monopole effects in supersymmetric theories in the presence of a large mass.

In this work we concentrate on vortices in $(2,1)$ spacetime dimensions, but we consider the case of a global configuration, which is known to have logarithmically divergent energy \cite{Col,VS,MS,V} in the infinitely large space. Because of this divergence, not too much interest has been given to Abelian global vortices nowadays. Here, however, we study the case of a complex scalar field that evolves under the global $U(1)$ symmetry, with the aim of finding a way to make the global vortex stable, engendering finite energy. The subject is of current interest and below we first show how to make the global vortex stable. The procedure leads us to the presence of scale invariance, so we take the opportunity and study another model, in which one adds a gauge field that responds to the presence of an electric charge in a medium with generalized permittivity. This possibility brings an interesting result, the presence of asymptotic freedom, which is known to be a feature of strong interactions \cite{N1,N2}. As shown below, in the passage from the first to the second model, we trade scale invariance for asymptotic freedom.  

We start the investigation with a single complex scalar field and use the Bogomol'nyi procedure \cite{bogo} to show how to describe stable vortex which is driven by the global $U(1)$ symmetry. We go further and consider another model, in which one trades scale invariance for the presence of asymptotic freedom, in an environment in the absence of non Abelian gauge symmetry. We then end the work summarizing the results and including new possibilities for future investigations.

\section{The Model}
Let us work in $(2,1)$ flat spacetime dimensions with the Lagrangian density
\be\label{lmodel}
	\LL = \partial_\mu\overline{\vphi}\,\partial^\mu\vphi- V(|\vphi|),
\ee
where the overline denotes the complex conjugation, $\vphi$ is a complex scalar field and $V(|\vphi|)$ is the potential. Here, we use the metric tensor $\eta_{\mu\nu}=(1,-1,-1)$ and take $\hbar=c=1$. The equation of motion associated to the Lagrangian density \eqref{lmodel} is
\be\label{eom}
\partial_\mu \partial^\mu \vphi + \frac{\vphi}{2|\vphi|}V_{|\vphi|} =0,
\ee
where $V_{|\vphi|} = \partial V/\partial|\vphi|$. To search for solutions that provide a global vortex, the standard procedure considers static configurations and the ansatz
\be\label{ansatz}
\vphi(r,\theta) = g(r)e^{in\theta},
\ee
where $n$ is an integer which stands for the vorticity of the field configuration. The function $g(r)$ obey the boundary conditions
\be\label{bc}
g(0) = 0 \quad\text{and}\quad g(\infty) = v.
\ee
Here, $v$ is a parameter involved in the symmetry breaking of the system and is supposed to appear in the potential to be chosen to specify the system under consideration. The equation of motion \eqref{eom} with the ansatz \eqref{ansatz} takes the form
\be\label{secansatz} 
\frac{1}{r} \left(r g^\prime\right)^\prime = \frac{n^2g}{r^2} + \frac12 V_{g},
\ee
with the prime denoting the derivative with respect to the radial coordinate. The energy density can be calculated in the usual manner. For the ansatz \eqref{ansatz}, it becomes
\be\label{rhoans}
\rho = {g^\prime}^2 + \frac{n^2g^2}{r^2} + V(g).
\ee
The equation of motion \eqref{secansatz} is of second order and usually present nonlinearities engendered by the potential. In order to get first order equations, we use the Bogomol'nyi procedure \cite{bogo} and introduce an auxiliar function $W(g)$ to write the energy density \eqref{rhoans} as
\be\label{rhobogo1}
	\rho = \left(g^\prime-\frac{W_g}{r}\right)^2 + V - \frac{1}{r^2}\left(W_g^2 - n^2g^2\right) +\frac{2}{r}\,W^\prime.
\ee
The above equation shows how the potential has to be written to attain the Bogomol'nyi bound. However, we notice that in order to define the model one needs to specify the auxiliary function $W$, so the procedure still leaves plenty of room for the construction of models of practical interest.

As the crucial step, we change $V\to V/r^2$, inspired by Ref.~\cite{prl}, which included similar modification in the potential. We do this and impose that the potential has the form
\be\label{valpha}
V(|\vphi|) = W_{|\vphi|}^2 - n^2|\vphi|^2,
\ee
where we included the second term to compensate for the the second term in the right hand side of Eq.~\eqref{rhoans}. In this case, the equation of motion \eqref{secansatz} becomes 
\be\label{eqmod}
\frac{1}{r} \left(r g^\prime\right)^\prime = \frac{1}{r^2}W_gW_{gg},
\ee
and the energy density \eqref{rhoans} changes to
\be\label{enedensi}
\rho = {g^\prime}^2 + \frac{1}{r^2}W_{g}^2.
\ee
Supposing that $g(r)$ is such that $W_g$ becomes constant asymptotically, there is no logarithmically divergent contribution to the energy anymore. This is different to the behavior shown in \cite{V} and also in \cite{Col,VS,MS}. In this scenario, we can integrate Eq.~\eqref{rhobogo1} and write
\be\label{bogo}
E= 2\pi\int_0^\infty rdr \left(g^\prime-\frac{W_g}{r}\right)^2 + E_B,
\ee
where $E_B$ identifies the Bogomol'nyi bound
\be\label{eb}
\begin{split}
E_B &= 4\pi \int dr\, W^\prime, \\
&= 4\pi\left|W(g(\infty))-W(g(0))\right|.
\end{split}
\ee
From Eq.~\eqref{bogo}, we see that the energy is bounded, i.e., $E\geq E_B$. If the solutions obey the first order equation
\be\label{fo}
g^\prime = \frac{W_g}{r},
\ee
the Bogomol'nyi bound is saturated and the energy is $E=E_B$. The first order equation \eqref{fo} is scale invariant and its solutions obey the equation of motion \eqref{eqmod} for the potential \eqref{valpha}. 

Notice that the procedure described above introduces the possibility to calculate the energy without knowing the solution explicitly, in a way similar to the standard Maxwell-Higgs model investigated in Ref.~\cite{bogo}; see also Ref.~\cite{NO}. Another interesting feature is that the global vortex model supports minimum energy configurations for several potentials, defined by the function $W(|\vphi|)$, differently from the standard Maxwell-Higgs model, which only admits the Bogomol'nyi procedure for a specific potential of the Higgs type.

The change $V\to V/r^2$ which we used to write \eqref{valpha} is inspired by \cite{prl}, where we added to the potential a space-dependent contribution that breaks translacional invariance; however, it contributed to evade the Derrick-Hobart scaling theorem and allowed the presence of localized structures in arbitrary spatial dimensions. In the case of planar structures, the space-dependent factor that we have to include in the potential is exactly $1/r^2$, as we are using in the current work. As we have just seem, it appears very naturally when one requires that the energy is minimized to the Bogomol'nyi bound $E_B$. There, in Ref.~\cite{prl} the spatial factor in the potential appeared to circumvent the Derrick-Hobart scaling theorem \cite{Der,Hob}; here, in the current work the same spatial factor appears to make the solution obey the Bogomol'nyi bound, giving rise to the first order equation \eqref{fo}. 

This type of modification is of current interest and has been used in different contexts, for instance, in the recent Refs.~\cite{A,B,C,D,E,mono}. In particular, in \cite{A,B,C} planar structures generated from real scalar field are used to model the presence of skyrmions in planar magnetic materials. Also, in \cite{D} the behavior of fermions in the background of a planar structure described by a real scalar field is investigated, and more recently yet, in \cite{E} we study the formation of vortices with internal structure, due to the coupling of the Maxwell-Higgs fields with local $U(1)$ symmetry and a neutral scalar field with global $Z_2$ symmetry, which contains a potential with a spatial dependence of the same form suggested above. Moreover, in Ref.~\cite{mono} we considered a non Abelian model in three spatial dimensions with symmetry $SU(2)\times Z_2$, to deal with magnetic monopoles with internal structure.

The presence of the negative term in the potential \eqref{valpha} is motivated by the vorticity. This is similar to what happens with Q-balls, in which the angular frequency contributes and gives rise to an effective potential \cite{coleman} which allows for the presence of stable solutions; see also Ref.~\cite{qball} and references therein, for other investigations on stable Q-balls. In the current case, the function $W_g$ works to spontaneously break the $U(1)$ symmetry of the system, in a way such that the term $W_g^2$ plays the role of an effective potential for the model, in a manner similar to the effective potential that appears to stabilize Q-balls. There is a restriction for the choice of the function $W(g)$: it has to allow the solutions to be compatible with the boundary conditions in Eq.~\eqref{bc}, which requires $W_g(g)$ to vanish at $g=0$ and $g=v$.

Although the Bogomol'nyi bound is attained, the fact that the potencial in Eq.~\eqref{valpha} may be negative in the region where the solution exists suggests that we further study the stability of the system. For this reason, let us investigate stability against small fluctuations of the static solution. The perturbed field has the form
\be
\phi(r,\theta,t) = g(r)e^{in\theta} + \eta(r,\theta,t),
\ee
in which $g(r)$ is a solution of Eq.~\eqref{secansatz}. Notice that, at this point, we are not yet considering the Bogomol'nyi procedure. By substituting the above expression in the time-dependent equation of motion \eqref{eom}, we get
\be\label{eqeta}
\partial_\mu\partial^\mu \eta + \frac1{4r^2}\left( \frac{V_g}{g} \left(\eta - e^{2in\theta}\,\overline{\eta}\right) + V_{gg}\left(\eta + e^{2in\theta}\,\overline{\eta}\right)\right) = 0.
\ee
The above equation admits a separation of variables in the form 
\be
\eta(r,\theta,t) = \sum_k \xi_k(r) e^{in\theta}\cos(\omega_k t).
\ee
By substituting this into Eq.~\eqref{eqeta}, we get
\be
-\frac1r\left(r\xi_k^\prime\right)^\prime + \frac1{2r^2}V_{gg}\xi_k + \frac{n^2}{r^2}\xi_k = \omega_k^2\xi_k.
\ee
If we consider the potential in Eq.~\eqref{valpha}, the first order equation \eqref{fo} is valid and the Bogomol'nyi bound is attained. In this situation, the above stability equation becomes
\be
-\frac1r\left(r\xi_k^\prime\right)^\prime + \frac{W_{gg}^2 + W_gW_{ggg}}{r^2}\xi_k = \omega_k^2\xi_k.
\ee
It may be written in the form $L \xi_k = \omega_k^2\xi_k$, with the Sturm-Liouville operator
$L$ factorized as $L=S^\dag\,S$, in the form \cite{SL}
\be
L = \left(\frac{d}{dr} + \frac{W_{gg}}{r}+\frac1{r}\right)\left(-\frac{d}{dr} + \frac{W_{gg}}{r}\right).
\ee
This ensures that there are only non-negative eigenvalues. Therefore, the global vortex solution is stable under small radial fluctuations.

\begin{figure}[t!]
\centering
\includegraphics[width=4.2cm]{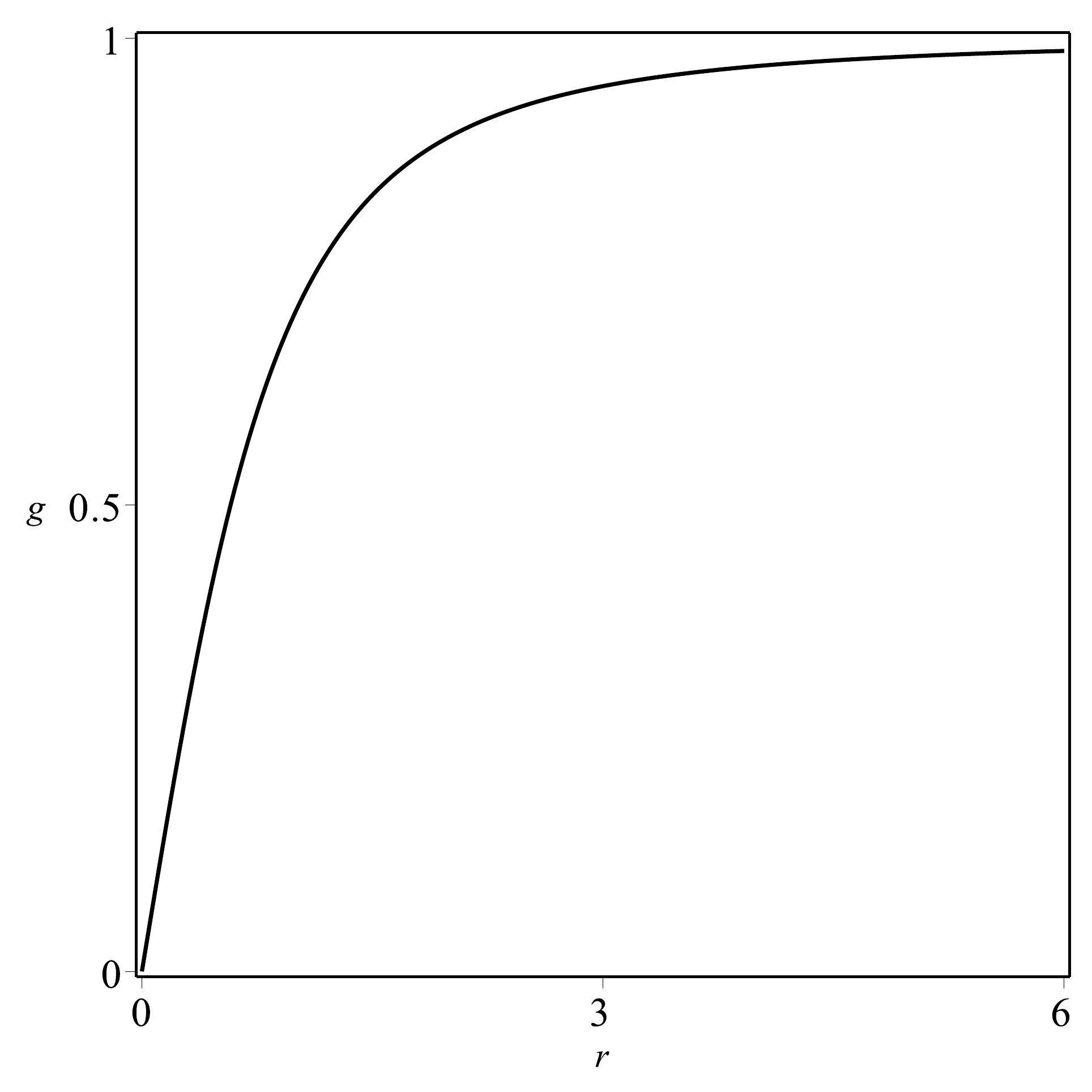}
\includegraphics[width=4.2cm]{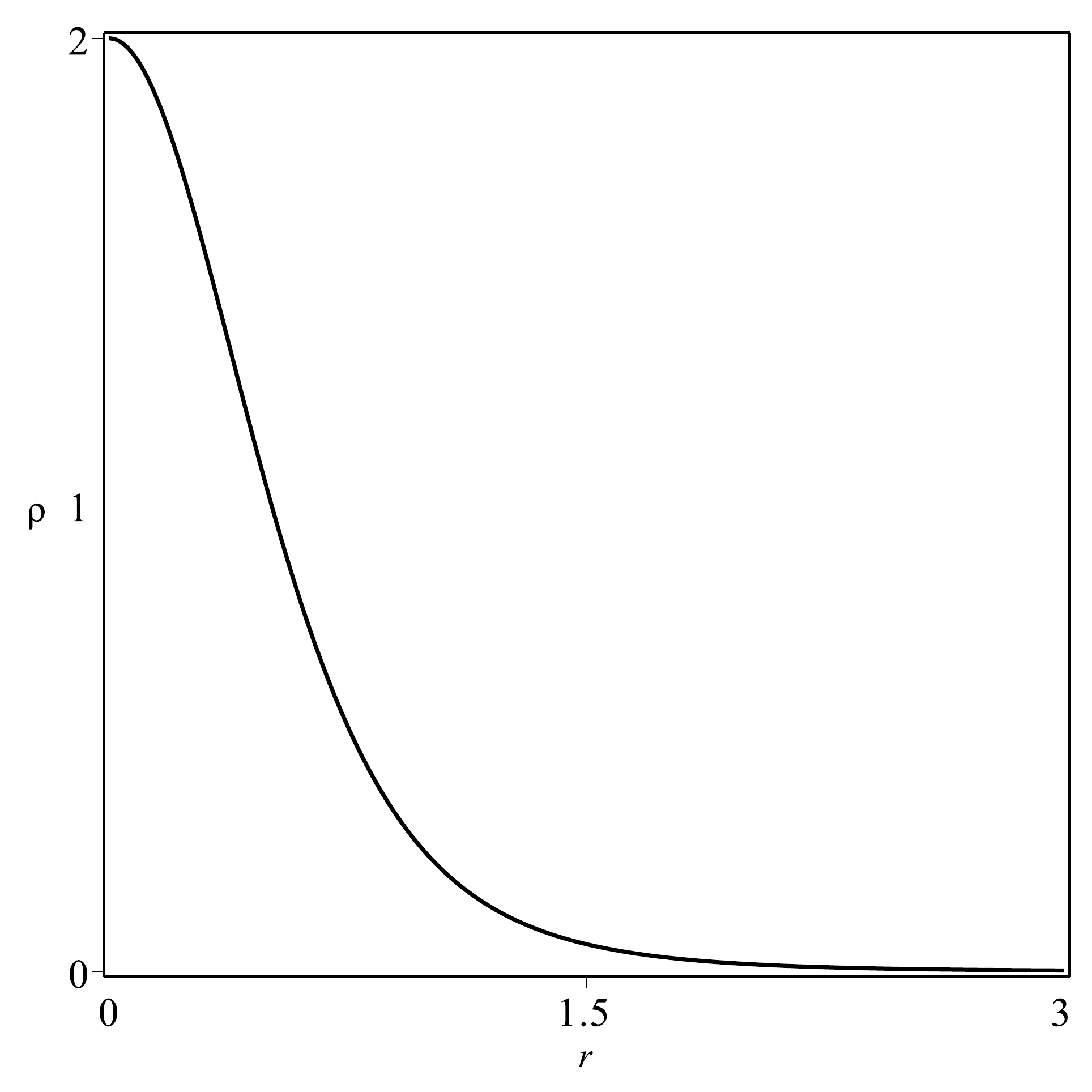}
\caption{The solution in Eq.~\eqref{g1} (left) and its energy density \eqref{rho1} (right). Here we take $r_0=1$.}
\label{fig1}
\end{figure}

We now provide an example of model described by the potential \eqref{valpha} and a specific function $W(|\vphi|)$. For simplicity, we consider dimensionless field. We then take
\be\label{wphi6}
W(|\vphi|) = \frac12|\vphi|^2 - \frac14|\vphi|^4.
\ee
This choice leads $W_{|\vphi|}$ to present minima at $|\vphi|=0$ and $|\vphi|=1$, matching with the boundary conditions \eqref{bc} for $v=1$. The first order equation \eqref{fo} becomes
\be
g^\prime = \frac{g(1-g^2)}{r}.
\ee
The solution can be obtained analytically; it is
\be\label{g1}
g(r) =\frac{r}{\sqrt{r_0^2 + r^2}},
\ee
where $r_0$ is an arbitrary parameter. The energy density \eqref{enedensi} takes the form
\be\label{rho1}
\rho(r) = \frac{2\, r_0^4}{(r_0^2 + r^2)^3}.
\ee
In Fig.~\ref{fig1}, we depict the solution \eqref{g1} and its energy density \eqref{rho1}, for $r_0=1$.

We see that the solution connects the minima $g=0$ and $g=1$ of the function $W$ for the choice in Eq.~\eqref{wphi6}. The energy density is smooth and vanishes at infinity. By integrating it, we get the value $E=\pi$, which matches with the value obtained through Eq.~\eqref{eb}. We emphasize that the model engenders scale invariance, and this is the price we had to pay: we found stable finite energy global vortices, but we cannot decide on its size. However, we can  circumvent scale invariance by considering a modified model which we investigate below.

\section{Presence of electric charge}
Let us now go further and trade scale invariance with an electric change in a medium with generalized permeability, which is used to capture the basic feature of asymptotic freedom. We modify the above model and work with a Lagrangian density that couples the complex scalar field to an Abelian gauge field which is engendered by an external source
\be\label{lmodelapp}
	\LL = \partial_\mu\overline{\vphi}\partial^\mu\vphi - \frac{P(|\vphi|)}{4}F_{\mu\nu}F^{\mu\nu} - A_\mu j^\mu,
\ee
where $A_\mu$ is the gauge field and $F_{\mu\nu}$ is the electromagnetic strength tensor. $P(|\vphi|)$ is included to describe a generalized medium and the current is $j^\mu = (e\delta(r),0,0)$, representing a point charge $e$ centered at the origin. This is different from other models which have been studied before; see, e.g., Ref.~\cite{BB} and references therein. 

The equations of motion associated to the Lagrangian density \eqref{lmodelapp} are
\bes\label{eomapp}
\bal\label{eomgapp}
\partial_\mu \partial^\mu \vphi + \frac{\vphi}{8|\vphi|}P_{|\vphi|}F_{\mu\nu}F^{\mu\nu} =0, \\ \label{meqs}
\partial_\mu\left(PF^{\mu\nu}\right) = j^\nu,
\eal
\ees
where $P_{|\vphi|} = \partial P/\partial|\vphi|$. The electric field is $E^i=F^{i0}= (E_x,E_y)$. Since $j^i=0$, there is no magnetic field. By setting $\nu=0$ and investigating static configurations, we get from Eq.~\eqref{meqs} that
\be\label{eapp}
{\bf E} = \frac{e}{rP(|\vphi|)} \hat{r},
\ee
where $\hat{r}$ is the radial unit vector. In the above equation, we see that the electric field may be modified by the function $P(|\vphi|)$, which can also be seen as a generalized electric permittivity. Since in the static case ${\bf E} = -\nabla A_0$, one may also show that the surviving component of the gauge field is
\be
A_0 = -e\int\frac{dr}{rP(|\vphi|)}.
\ee
The case $P=1$ recovers Coulomb's law in the plane, with ${\bf E} = e\,\hat{r}/r$ and $A_0 = -e\ln (r)$. Regarding the equation of motion \eqref{eomgapp}, we can substitute the ansatz \eqref{ansatz} to get
\be\label{gapp}
\frac{1}{r} \left(r g^\prime\right)^\prime = \frac{n^2g}{r^2} - \frac{P_g}{4}|{\bf E}|^2.
\ee
Combining it with Eq.~\eqref{eapp}, we can write
\be
\frac{1}{r} \left(r g^\prime\right)^\prime = \frac{1}{2r^2}\frac{d}{dg}\left(n^2g^2+ \frac{e^2}{2P}\right).
\ee
We suppose that 
\be
P(|\vphi|) = \frac{e^2}{2\left(W_{|\vphi|}^2 - n^2|\vphi|^2\right)},
\ee
to write the equation of motion \eqref{gapp} as in Eq.~\eqref{eqmod}. Therefore, the generalized electric permittivity appears to be controlled by the global vortex which we discussed before. The energy of the configuration is
\be
\rho = {g^\prime}^2 + \frac{n^2g^2}{r^2} +\frac12P(g)E^2 + A_0j^0
\ee
or better
\be
\rho= {g^\prime}^2 + \frac{W_g^2}{r^2}  + \rho_e,
\ee
where $\rho_e = e A_0 \delta(r)$. 

In this case, we can write the energy as $E = E_{vortex} + E_e$, where $E_{vortex}$ is given by Eq.~\eqref{bogo} and $E_e = 2\pi e \lim_{r\to0}(rA_0)$ is the energy associated to the point charge $e$ in the medium where the generalized electric permittivity is described by the function $P(g)$. If $E_e=0$ and Eq.~\eqref{fo} holds, the energy is minimized to $E=E_B$.

\begin{figure}[t!]
\centering
\includegraphics[width=6cm]{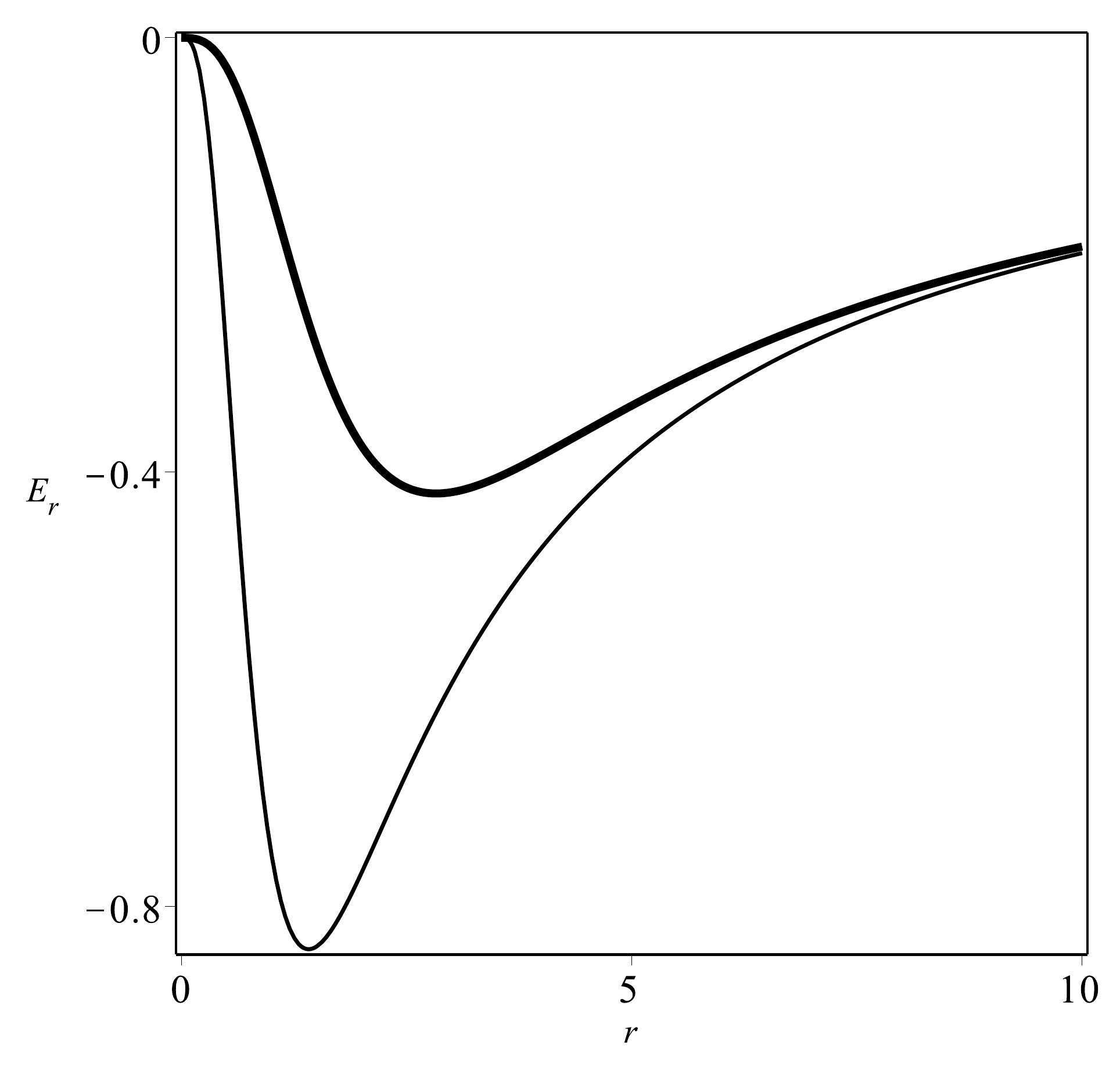}
\caption{The electric field in Eq.~\eqref{eapp1}, depicted for $n=1$ and for $r_0=1$ and $2$, with the thickness of the lines increasing with $r_0$.}
\label{fig2}
\end{figure}

\begin{figure}[h!]
\centering
\includegraphics[width=3.8cm]{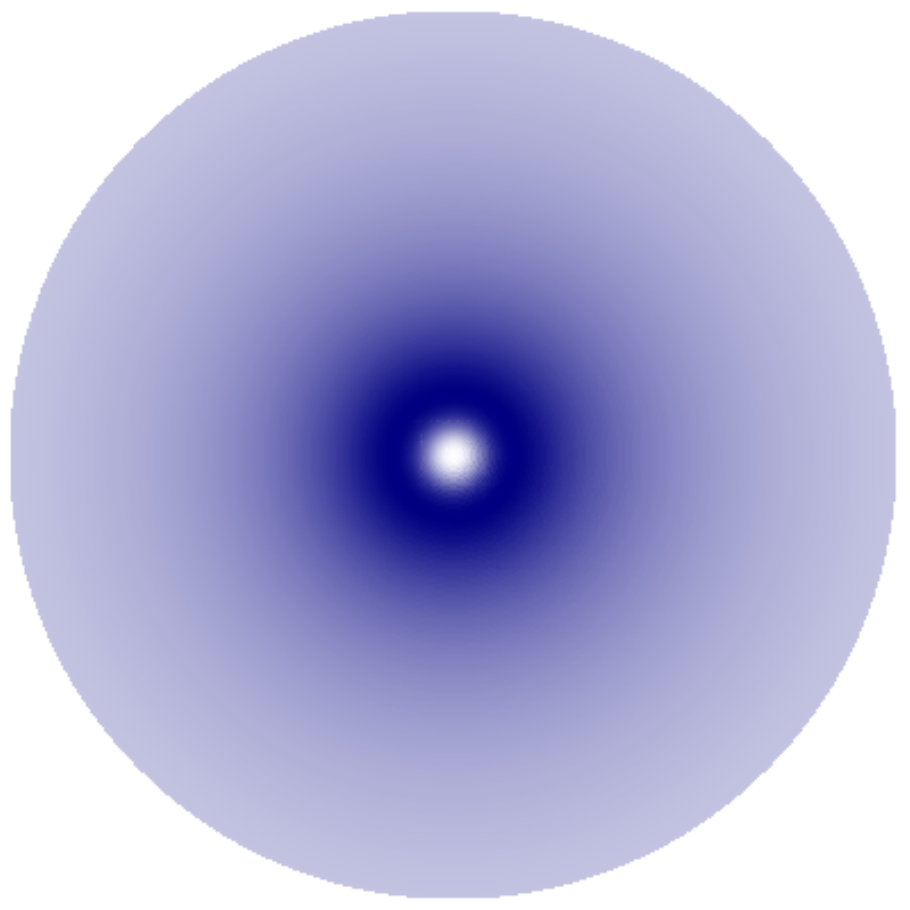}\hspace{0.4cm}
\includegraphics[width=3.8cm]{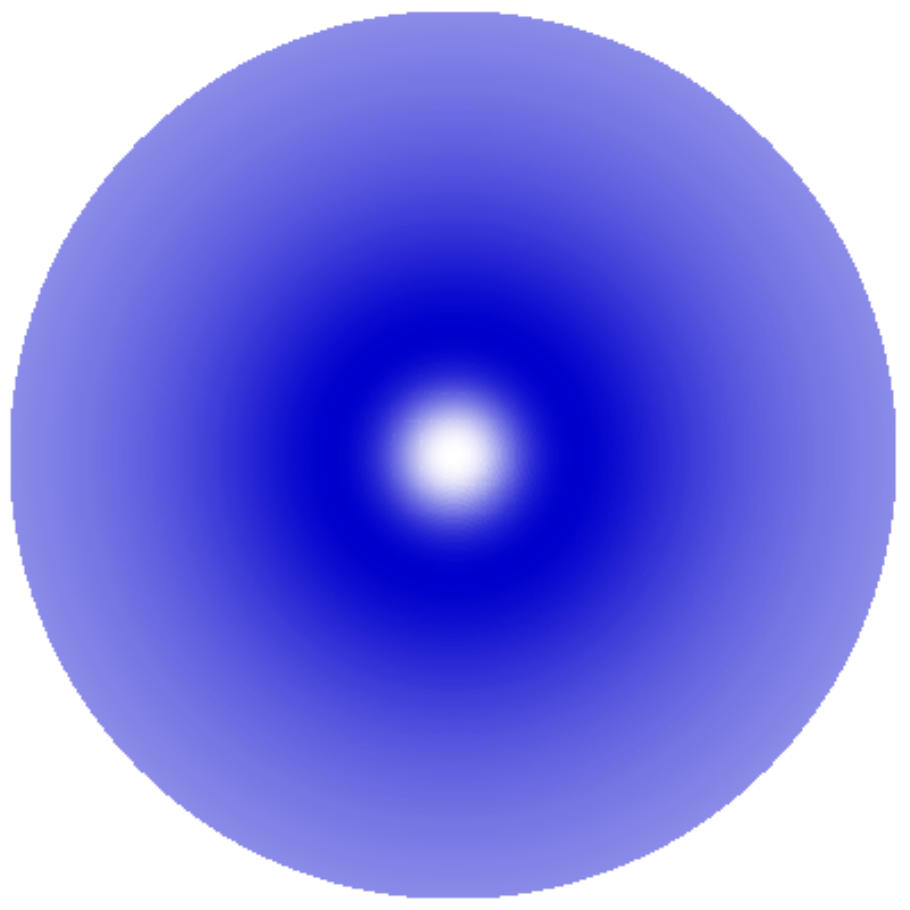}
\caption{The electric field in Eq.~\eqref{eapp1} depicted in the plane for $n=1$ and $r_0=1$ (left) and $2$ (right), with the darkness of the color related to the increasing of its intensity.}
\label{fig3}
\end{figure}

We can calculate the electric field for $e=1$ and $W(g)$ as in Eq.~\eqref{wphi6}. The result is analytical, and can be expressed in the form 
\be\label{eapp1}
{\bf E} = -\frac{2\,r}{r_0^2+r^2}\left(n^2-\frac{r_0^4}{(r_0^2+r^2)^2}\right) \hat{r}.
\ee
The electric field can be seen in Fig.~\ref{fig2}. It has an interesting behavior, which motivated us to depict it in the plane in Fig.~\ref{fig3}.

We see that it extends far from the core of the vortex and differs from the Coulomb interaction, since it vanishes at $r=0$, where the charge is placed. The above equation shows an interesting feature: a single positive charge generates a convergent electric field, pointing in its direction. This is the opposite behavior of the particle in the vacuum and it happens because of the presence of the generalized permittivity $P(|\vphi|)$. In this case, one can also show that
\be
A_0 = \frac{r_0^4}{2\left(r_0^2+r^2\right)^2} + n^2 \ln(r_0^2+r^2),
\ee
which can be used to show that $E_e=0$. Then, the energy is given by Eq.~\eqref{eb}, which leads to $E = \pi$. Therefore, even though the permittivity is negative, the energy of the system is positive. This behavior also regularizes the energy of the charge placed at the origin. 

The unusual behavior of the electric field is very interesting, since it simulates asymptotic freedom \cite{N1,N2}. In
Fig.~\ref{fig3} the white color that appears around the origin identifies a region with vanishing electric field. In this sense, we studied a simple Abelian model that captures the essence of the asymptotic freedom, and now the in principle arbitrary parameter $r_0$ can be estimated experimentally.\\

\section{Ending comments}
In this work we introduced a relativistic model that supports stable finite energy global static configurations in $(2,1)$ spacetime dimensions. Such global vortices admit a first order formalism in which the Bogomol'nyi bound can be attained. The vortex solution and its energy density are obtained analytically, and the price we had to pay for stability of the global configuration is scale invariance, that is, the model we constructed obeys scale invariance, so one cannot decide on the size of the topological structure. 

To circumvent scale invariance, we changed the model, trading the potential of the complex scalar field for the presence of an electric charge in a medium with generalized permittivity. In this new scenario, some interesting features have arisen. In particular, the electric field regularized the energy of the charge at the origin and pointed toward a positive charge, which is the opposite behavior described by the standard Coulomb's law. Although we investigated a simple Abelian model, it correctly captured the asymptotic freedom behavior, since the electric field vanishes as one approaches the electric charge placed at the origin. As the size of the asymptotic region depends on $r_0$, we can now estimate $r_0$ experimentally.

The presence of stable global vortices which obey first order differential equation suggests that the model can be the bosonic portion of a larger, supersymmetric theory. This is an interesting issue which deserves further investigations, to see how supersymmetry is working in this new environment. The case of two complex scalars is also of interest, with the global symmetry $U(1)$ extended to be $U(1)\times U(1)$, to account for the interaction of visible and hidden sectors \cite{VH} via coupling between the two scalar fields, known as the Higgs portal \cite{HP}. Other possibilities include the study of nonlinear sigma models and the case of nonrelativistic field theories with direct applications to systems of condensed matter. In particular, the Gross-Pitaevskii equation \cite{GP} can be investigated under similar conditions, and since it provides an appropriate description of Bose-Einstein condensates at the mean-field level for ultra-cold temperatures, it may offer a novel route to investigate global vortices in condensates. 

\acknowledgments{We would like to acknowledge the Brazilian agency CNPq for partial financial support. DB thanks support from grant 306614/2014-6, MAM thanks support from grant 140735/2015-1 and RM thanks support from grant 306826/2015-1.}

\end{document}